\documentclass[preprint,letterpaper,showpacs]{revtex4}
\def\be{\begin{equation}}
\def\ee{\end{equation}}
\def\bear{\begin{eqnarray}}
\def\eear{\end{eqnarray}}
\def\bearst{\begin{eqnarray*}}
\def\eearst{\end{eqnarray*}}
\usepackage{epsfig}

\begin{document}

\title{Analytic determination of the asymptotic quasi-normal mode
spectrum of small Schwarzschild-de Sitter black holes} 

\author{K. H. C. Castello-Branco}

\email{karlucio@fma.if.usp.br} 

\author{E. Abdalla}

\email{eabdalla@fma.if.usp.br}

\affiliation{Universidade de S\~ao Paulo, Instituto de F\'\i sica\\
Caixa Postal 66318, 05315-970, S\~ao Paulo-SP, Brazil.}

\begin{abstract}

Following the monodromy technique performed by Motl and Neitzke, we
consider the analytic determination of the highly damped (asymptotic)
quasi-normal modes of {\it small} Schwarzschild-de Sitter (SdS) black holes. 
We comment the result as compared to the recent numerical data of
Konoplya and Zhidenko.
\end{abstract}


\maketitle

For intermediate times, perturbations of a black hole are dominated
by characteristic, damped oscillations - the so-called 
quasi-normal modes (QNMs), the pure tons it emitts in 
the final stage of its perturbation (for reviews, 
see \cite{kokkotas&schmidt-nollert-rev}). 

The search for large 
imaginary, highly damped QNM frequencies has 
recently attracted a great deal of interest. This has been motivated by the
work of Hod, who, based on the Bohr correspondence principle, proposed that
the real part of the asymptotic, highly damped QNM frequencies of a 
black hole can be used to determine the spacing (fundamental unit) of 
its quantum area 
spectrum \cite{hod}. The quantization of black holes horizon area 
has been proposed by Bekenstein thirty years ago
\cite{Bek-quant-ar}, based on 
the adiabatic invariance of the horizon area and the analogy with the 
quantum mechanics of adiabatic systems (for a review, see 
\cite{bek-escola}). In \cite{hod}, Hod has used 
the highly damped spectrum of QNMs of a Schwarzschild black hole, 
numerically obtained 
by Nollert \cite{nollert-assnt} and later confirmed by Andersson 
\cite{anders} and Liu \cite{liu}. Recently, Dreyer \cite{dreyer} used 
Hod's conjecture in the context of Loop Quantum Gravity to fix a 
free-parameter (the Barbero-Immirzi parameter) which appears in the
area spectrum provided by this theory (for reviews, see 
\cite{rovelli-smolin}). In this way,
he could naturally obtain the Bekenstein-Hawking entropy (see also 
\cite{corichi}). Thereafter, several works concerning the highly damped QNMs 
and their role in the black hole area quantization have appeared 
\cite{varios}(for a review and more references, see \cite{carlemyosh-review}). There have also been studies along the same 
lines, but concerning non-asymptotically flat black holes, namely, 
black holes in asymptotically de Sitter (dS) and anti-de Sitter (AdS) 
space-times \cite{varios-ds-ads}. Furthermore, the study of QNMs of
dS \cite{varios-ds} and AdS \cite{varios-ads} black holes has been
under intensive study. Finally, we mention that the works of Motl 
\cite{motl} and Motl and Neitzke \cite{m&n} deserve special 
attention, since they first provided 
an analytic determination of the large imaginary QNMs of a  
Schwarzschild black hole, in agreement with the numerical formula of 
Nollert \cite{nollert-assnt}. Recently, following the approach of Motl and
Neitzke in \cite{m&n}, Krasnov and Solodukhin 
\cite{kras-solod} rederived the real part of the highly damped QNMs of a 
Schwarzschild black hole by means of a more detailed study, on the
complex plane, of the relationship between the monodromy associated to
the type of the differential equation which appears in black hole 
perturbation problem and the related Riemann surfaces.

The Schwarzschild-de Sitter (SdS) black hole metric in Schwarzschild 
coordinates $(t,r,\theta,\phi)\,$ is given by \cite{gibhawk} (we will be using
natural units, $G=c=k=\hbar=1\,$, throughout the text) 
$ds^{2}= -f(r)dt^{2} + f(r)^{-1}dr^{2} + r^{2}(d\theta^{2} + \sin^{2}\theta
d\phi^{2})\,$, where $f(r)= 1-\frac{2M}{r}-\frac{r^2}{a^2}\,$. $M$ is 
the black hole mass and $a$ the ``cosmological radius``, related to 
the cosmological 
constant $\Lambda > 0$ by $a^2=3/\Lambda$. Describing a black hole in 
a cosmological background, $M$ has to be limited to the interval 
$0 < M < M_{N}=a/3\sqrt{3}\,$. In this interval, the 
function $f(r)$ has three 
distinct real zeros : at $r=r_e\,$, the black hole event horizon, 
$r_c\,(\,> r_e)\,$, the cosmological event horizon, and 
$r_0 = -(r_e + r_c)\,$. The extreme 
value $M_{N}\,$ gives the Nariai limit \cite{nariai}, for which 
the event and cosmological horizons coincide. 

Associated to each horizon $r_i\,$ there is a surface gravity 
$\kappa _{i}\,$, defined by 
$\kappa _{i}=\frac{1}{2}|\frac{df}{dr}|_{r=r_i}\,$. Explicitly, we 
have $\kappa_e = (r_c-r_e)(r_e-r_0)/2a^2r_e\,$, $\kappa_c =
(r_c-r_e)(r_c-r_0)/2a^2r_c\,$, and 
$\kappa_0 =(r_e-r_0)(r_c-r_0)/2a^2(-r_0)\,$. 
The metric parameters $M$ and $a$ can be written in terms of $r_e$ and $r_c$ as
$a^2= r_e^2 + r_c^2 + r_e r_c$ and $2Ma^2= r_e r_c(r_e +r_c)\,$. 

Assuming a time dependence $e^{i\omega t}\,$, expanding the 
perturbating field into multipoles and defining the 
tortoise coordinate $x$ by $dx/dr = 1/f(r)\,$, one can show that 
the equations for scalar, gravitational and electromagnetic perturbations 
of a SdS black hole are summarized by \cite{mellor-moss-guven-nunez}
\be
\frac{d^{2}\psi(r)}{dx^{2}} +
\left [\omega^{2}-V(r(x))\right ]\psi(r) = 0 \quad, 
\label{eqondaunid}
\ee
where the potential $V(r)$ is 
\begin{equation}
V(r)=f(r)\left[\frac{\ell(\ell+1)}{r^{2}}-\frac{2M(j^{2}-1)}{r^{3}}
\right]\quad,
\label{pots-ds}
\end{equation}
for gravitational ($j=2$) and electromagnetic ($j=1$) perturbations. 
$\ell$ and $j$ are the multipole index and the spin of the 
perturbating field, respectively. We will consider here only
electromagnetic and axial 
gravitational perturbations (in four dimensions, 
axial and polar gravitational perturbations are isospectral). 

Explicitly, the tortoise coordinate is given by 
\be
x=\frac{1}{2\kappa_e}\ln\left (\frac{r}{r_e}-1\right )-\frac{1}{2\kappa_c}
\ln\left (1-\frac{r}{r_c}\right )+\frac{1}{2\kappa_0}\ln\left (1-\frac{r}{r_0}\right)\,.
\label{tarta-sds}
\ee

Since $V(x)\equiv V(r(x))$ goes to zero exponentially for
$x\rightarrow\pm\infty\,$, in terms of the tortoise coordinate 
the boundary conditions for the QNMs of a SdS black hole are typically 
the same as those of an asymptotically Schwarzschild black hole, namely,
ingoing waves at the black hole horizon ($x\rightarrow -\infty$) and
outgoing waves at the cosmological horizon 
($x\rightarrow +\infty$), {\it i.e.}, in view of the dependence
$e^{i\omega t}\,$ we have 
\begin{equation}
\psi\sim e^{\mp i\omega x}\qquad,\quad x\rightarrow\pm\infty\quad.
\label{compass-ds}
\end{equation}

Motl and Neitzke \cite{m&n} have considered the analytic extension to
the complex $r$-plane of the perturbation equation (similar to 
(\ref{eqondaunid})) for the Schwarzschild metric. 
Since in this case the equation has (regular) singular points 
at the origin ($r=0$) and at the black hole horizon ($r=r_e\,$), its 
solution $\psi(r)$ is multivalued around these points. Such  
multivaluedness is a crucial feature in the analysis 
of \cite{m&n}, for the key 
idea is to use the QNM boundary conditions to compute the monodromy 
of $\psi(r)\,$ as we run along a conveniently 
chosen contour in the $r$-plane. Comparison of 
the monodromy calculated locally 
(by solving the perturbation equation directly) 
and globally (by direct application of the QNM boundary conditions)
allowed the determination of the highly damped 
quasi-normal modes, {\it i.e.}, those with $Im\,\omega >\,> Re\,\omega\,$. 

Although the physical region of interest for the problem of QNMs of a
SdS black hole is $r_e < r <r_c\,$, an analytic extension to the
region $0 < r < r_c$ is essential for a ``monodromic analysis``. As 
we will verify later, the Schwarzschild and SdS cases have similar
features which concern the behaviour of the potential near the black
hole singularity ($r=0\,$).

Due to the logarithmic terms in (\ref{tarta-sds}), the analytic 
continuation to complex-$r$ implies that $x$ is a multivalued function
of $r\,$. But $Re\,x\,$ is free of ambiguity and we can determine 
its sign in the $r$-plane. The multivaluedness of $x(r)$ is also 
fundamental in this approach. Together with the multivaluedness 
of $\psi(r)$ around the (regular) singular point $r_e\,$, it
will lead us to the results we look for. 

As in \cite{m&n}, we also conveniently introduce the 
variable $z = x + const\,$. Choosing the constant such that $z=0$ 
for $r=0\,$, and the branch $n=0$ for $\ln(\pm 1)\,$, from 
(\ref{tarta-sds}) we have $z= x - \pi i/2\kappa_e\,$.

By analytically continuing $r$ we can distinguish the functions 
$e^{\mp i\omega x}$ by their monodromy at $r=r_{e}\,$. This enables 
us to define the boundary condition at $r=r_e$ by simply requiring that 
$\psi(r)$ have monodromy $e^{2\pi(\omega/2\kappa_{e})}=e^{\pi\omega/\kappa_e}$ 
on a clockwise contour around this point. We note that we cannot do
the same at $r=r_c\,$, since as $r$
is now complex, $x$ does not tend to $+\infty$ for
$r=r_c\,$. Nevertheless, if we consider the special case of 
{\it small} black holes ($r_c > > r_e\,$), we can approximately 
consider that $x >> 1\,$ for $r=r_c\,$, 
as an expansion of $x(r)$ for $(r_c/r_e) >> 1$ shows.  

\begin{figure}[hbt]
\begin{center}
\mbox{\epsfig{file=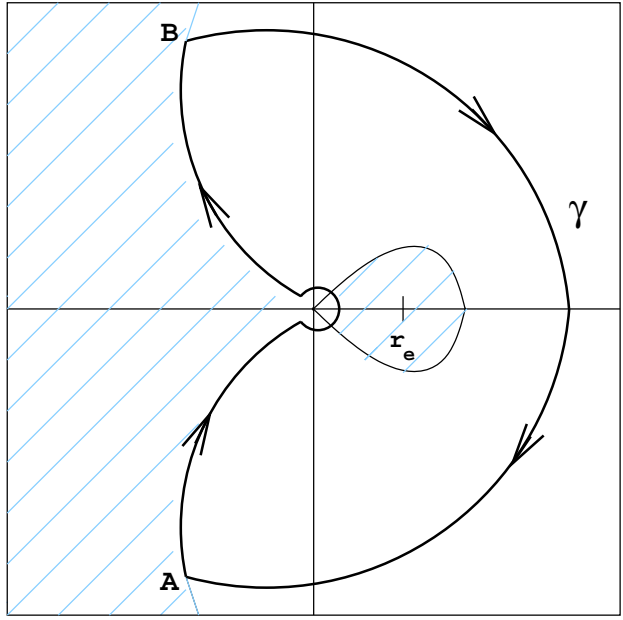, width=0.5\textwidth}}
\end{center}
\caption{The complex $r$-plane and the contour $\gamma\,$. The hachured 
part marks the region where $Re\,x\,<\,0\,$.}
\label{fig1}
\end{figure}

Now we can closely follow \cite{m&n} and in order to define 
the boundary condition at $r_c\,$, we simply analytically continue $\psi(r)$ via rotation to
the line $Im\,(\omega x)=0\,$. Hence we will have a purely oscillatory
asymptotic behaviour on the real line $\omega x\,$, what implies that 
it is possible to select a particular solution by specifying its 
asymptotics. We will match the asymptotics along the line $Re\,x=0$ 
(corresponding to the left part of the contour $\gamma$ between $A$ and
$B$ in the figure \ref{fig1}) 
and use the boundary conditions to determine the QNMs. 
Taking the limit where $\omega$ is almost purely imaginary (highly
damped modes), we see that the line $Im\,(\omega x)=0$ is slightly 
sloped off the line $Re\,x=0\,$. Since there are two possible 
directions for the rotation, we can choose the one corresponding 
to an angle smaller than $\pi/2\,$.

From (\ref{compass-ds}), we note that for QNMs we have 
$Im\,\omega\,>\,0\,$. Assuming initially that 
$Re\,\omega\,>\,0\,$, in view of the condition 
$Im\,(\omega x)=0\,$, we see that $x=+\infty$ is rotated to 
$\omega x= +\infty\,$, and on this line the 
boundary condition at $x=+\infty$ is 
\be
\psi(r)\sim\,e^{-i\omega x}\quad, \qquad \omega x\rightarrow +\infty\quad. 
\label{condcont-infto}
\ee

Let us first compute the monodromy locally. This will be done by 
matching the asymptotics along the line $Im\,(\omega x)=0$ (or 
along $Im\,(\omega z)=0\,$, for $\omega$ (almost) purely imaginary). 
We will start at $A$ and move along the contour $\gamma$ towards the 
origin (see figure \ref{fig1}). The asymptotics at $A$ can be matched to that 
near the origin. First, to obtain the solution near the origin, we expand 
$x$ near $r=0\,$ and thus obtain for small $z$     
\be
z\approx \beta r^{2}\quad,
\label{zproxde0}
\ee
where
$\beta = \frac{1}{4}\left
(\frac{1}{\kappa_{c}r_{c}^{2}}-\frac{1}{\kappa_{e}r_{e}^{2}}-\frac{1}{\kappa_{0}r_{0}^{2}}\right)\,$.

Therefore, near the origin the potential behaves as 
\begin{equation}
V(z)\approx\frac{4M^{2}\beta^{2}(j^{2}-1)}{z^{2}}\quad.
\label{pot-orig-sds}
\end{equation} 

We remark that (\ref{zproxde0}) has been deduced without using the
approximation of small black holes. Hence, the near origin behaviour 
(\ref{pot-orig-sds}) remains valid in general. 

In view of (\ref{pot-orig-sds}), (\ref{eqondaunid}) can be reduced to a 
Bessel equation. Its solution near the origin is then given by  
\be
\psi(z)=A_{+}c_{+}\sqrt{\omega z}J_{\nu}(\omega z) +
A_{-}c_{-}\sqrt{\omega z}J_{-\nu}(\omega z) \quad,
\label{sol-bessel}
\ee 
with the index $\nu$ being given by
\be
\nu=\frac{1}{2}\sqrt{1 + (4M\beta)^{2}(j^{2}-1)}\quad.
\label{index-bessel}
\ee

We will now match the solution (\ref{sol-bessel}) to the 
asymptotics away from the origin. 
Taking the asymptotic behaviour of $J_{\pm\nu}(\omega z)$ as 
$\omega z\rightarrow\infty\,$, we can choose the constants $c_{\pm}$ in
(\ref{sol-bessel}) such that 
\be
c_{\pm}\sqrt{\omega z}J_{\pm\nu}(\omega z)\,\sim\,2\cos (\omega z - \alpha
_{\pm})\,  \qquad\hbox{as}\quad\omega z\rightarrow\infty\quad,
\label{comp-asst-bessel}
\ee
where $\alpha _{\pm}=\frac{\pi}{4}(1 \pm 2\nu)\,$. 

From (\ref{comp-asst-bessel}) and (\ref{sol-bessel}), and making use of 
the boundary condition (\ref{condcont-infto}), we can write for the 
asymptotics at $A$ 
\be
\psi(z)\sim (A_{+}e^{i\alpha _{+}} + A_{-}e^{i\alpha _{-}})e^{-i\omega z}\,  
\qquad\hbox{as}\quad\omega z\rightarrow\infty\quad,
\label{psiA}
\ee
since
\be
A_{+}e^{-i\alpha _{+}} + A_{-}e^{-i\alpha _{-}}\,=\,0\quad.
\label{vinculo}
\ee

In order to follow the contour until $B$ we have to turn around the origin. 
Thus we must perform a rotation of $3\pi/2$ in the $r$-plane,
corresponding to $3\pi$ in the $z$-plane. Using the asymptotic
behaviour of the Bessel function near the origin, we can match it to 
the asymptotic solution. In fact, using for the Bessel function 
the behaviour 
$J_{\pm\nu}(z)\,\sim\,\left(\frac{z}{2}\right)^{\pm\nu}\,$ near the origin,
after the $3\pi$ rotation we have 
$c_{\pm}\sqrt{\omega z}J_{\pm\nu}(\omega z)\,\sim\,e^{6i\alpha _{\pm}}
2\cos(-\omega z - \alpha _{\pm})\,$, as $\omega z\rightarrow -\infty\,$,
which combined with (\ref{sol-bessel}) gives for the aymptotics at $B$
\be
\psi(z)\sim (A_{+}e^{5i\alpha _{+}} + A_{-}e^{5i\alpha _{-}})e^{-i\omega
z}+(A_{+}e^{7i\alpha _{+}} + A_{-}e^{7i\alpha _{-}})e^{i\omega z}\, \quad
\hbox{as}\quad\omega z\rightarrow -\infty\,.
\label{psiB}
\ee

We can go from $B$ back to $A$ along the large semi-circle 
($|r|\sim |r_{c}|$). In this region, taking into account our
approximation of small black holes, we have $x >> 1\,$, and thus 
$V(x)$ goes to zero, such that ${\omega}^2$ dominates the potential, 
and we can approximate the solution of the wave equation there by plane waves. 
Hence, the coefficient of $e^{-i\omega z}$ remains essentially 
unchanged as we return to $A\,$, whereas the same cannot be 
said for the coefficient of $e^{i\omega z}\,$, which gives an
exponentially small contribution to $\psi (r)$ in the region where 
$Re\,x\,>\,0\,$. Therefore, from (\ref{psiA}) and (\ref{psiB}), we 
see that the monodromy around the contour $\gamma$ is simply given by
\be
\frac{A_{+}e^{5i\alpha_{+}} + A_{-}e^{5i\alpha_{-}}}{A_{+}e^{i\alpha_{+}} + A_{-}e^{i\alpha_{-}}} = -(1+2\cos2\pi\nu)\quad,
\label{locmonodr}
\ee
where we have taken into account (\ref{vinculo}).

The global computation of the monodromy around $\gamma\,$ can be 
obtained by making use of the QNM boundary condition at $r_e\,$, since
the only singularity of $\psi(r)$ or $e^{-i\omega z}$ inside the contour 
$\gamma$ occurs at this point. The boundary condition at $r_e$ implies
that after a full trip around the contour clockwise, $\psi (r)$
acquires a phase $e^{\pi\omega/\kappa _e}\,$, while $e^{-i\omega z}$ 
acquires a phase $e^{-\pi\omega/\kappa _e}\,$. Therefore, we must 
multiply the coefficient of $e^{-i\omega z}$ in the asymptotics of 
$\psi(r)$ by $e^{2\pi\omega/\kappa _e}\,$. 

Comparing the local monodromy (\ref{locmonodr}) with the global 
one, we find 
\be
e^{2\pi\omega_n/\kappa_{e}}= -(1+2\cos2\pi\nu)\quad.
\label{exp-omega-final}
\ee

For the choice $Re\,\omega\,<\,0\,$ we would have to consider
the rotation in the opposite direction and then reverse 
considerations on the points $A$ and $B\,$, as well as to run the contour
$\gamma$ in the counter-clockwise direction. We thus conclude that 
the asymptotic, highly damped quasi-normal mode spectrum of {\it small} SdS
black holes is given by 
\be
\frac{\omega_n}{\kappa_e}= (n+\frac{1}{2})i\,\pm\,\frac{1}{2\pi}\ln\,|1 + 2\cos2\pi\nu|\quad,
\label{omega-assint-sds}     
\ee
with $n\rightarrow\infty\,$. 

From the relation $4M\beta = -\left[ 1 +\frac{r_e r_c}{2(r_e +
r_c)^{2}}\right]\,$, if we take the limit as $r_c\rightarrow\infty\,$,
which corresponds to $\Lambda=0\,$, we find $4M\beta = - 1\,$, such
that we recover the result of Motl and Neitzke for 
asymptotically flat Schwarzschild black holes \cite{m&n}, 
\be
8\pi M\omega_n = (2n+1)\pi i \pm \ln\,|1+2\cos\pi j|\quad .
\label{resul-lamb-zero} 
\ee

The result (\ref{omega-assint-sds}) shows that the real part of the
asymptotic modes has a constant value and then it fails to reproduce
the oscillatory behaviour recently found by Konoplya and Zhidenko 
\cite{konop-zid} in the case of gravitational perturbations ($j=2$). 
Nevertheless, for the electromagnetic case ($j=1\,$), $Re\,\omega_n =0\,$,
what agrees with the result of those authors, but the 
behaviour of the imaginary part of the modes, as given by 
(\ref{omega-assint-sds}), only partially agrees with that found in 
\cite{konop-zid}, since it does not show an oscillatory behaviour 
in its spacing, found for both gravitational and electromagnetic
cases in \cite{konop-zid}. Accorging to \cite{konop-zid}, an
oscillatory term should appear in $Im\,\omega_n\,$, in addition to 
$(n+\frac{1}{2})\kappa_e\,$.   

Summarizing, by applying the monodromy technique as performed by
Motl and Neitzke \cite{m&n}, we have considered the determination of the asymptotic, 
highly damped quasi-normal mode spectrum of {\it small} 
($r_e << r_c$) Schwarzschild-de Sitter black holes. However, we note 
that before we have restricted the analysis for small black holes, we verified that the dominant 
contribution for potential near the black hole singularity is 
exactly of the same type as found for the asymptotically
flat case, {\it i.e.}, the solution is a Bessel function around the 
origin. This result is important in applying the monodromy 
approach to find the asymptotic QNMs for general SdS black holes. 
Compared to the numerical results of Konoplya and Zhidenko 
\cite{konop-zid}, it seems our results do not give the correct 
asymptotic QNM frequencies for small SdS black holes. 
Consequently, in the context of the monodromy technique, it 
is important to search for the general analytic solution and 
compare it with the available numerical data in \cite{konop-zid}. 
In fact, this has already been done very recently by Cardoso {\it et
al.} and it will appear soon \cite{cardoso-priv-com}. 

\begin{acknowledgments}
The authors acknowledge V. Cardoso, R. Konoplya, C. Molina, and
A. Neitzke, for helpful discussions, and E. Asano for drawing the picture. This
work has been supported by the Brazilian agencies {\bf FAPESP} 
({\it Funda\c{c}\~ao de Amparo \`a Pesquisa no Estado de S\~ao Paulo})
and {\bf CNPq} ({\it Conselho Nacional de Desenvolvimento Cient\'\i fico e
Tecnol\'ogico}).
\end{acknowledgments}


\end{document}